**Wide-range epitaxial strain control of electrical and magnetic properties in high-quality $SrRuO_3$ films**

Yuki K. Wakabayashi,[1,*] Shingo Kaneta-Takada,[1] Yoshiharu Krockenberger,[1] Yoshitaka Taniyasu,[1] and Hideki Yamamoto[1]

[1]*NTT Basic Research Laboratories, NTT Corporation, Atsugi, Kanagawa 243-0198, Japan*

[†]Author to whom correspondence should be addressed: yuuki.wakabayashi.we@hco.ntt.co.jp

Abstract

Epitaxial strain in 4*d* ferromagnet $SrRuO_3$ films is directly linked to the physical properties through the strong coupling between lattices, electrons, and spins. It provides an excellent opportunity to tune the functionalities of $SrRuO_3$ in electronic and spintronic devices. However, a thorough understanding of the epitaxial strain effect in $SrRuO_3$ has remained elusive due to the lack of systematic studies. This study demonstrates wide-range epitaxial strain control of electrical and magnetic properties in high-quality $SrRuO_3$ films. The epitaxial strain was imposed by cubic or pseudocubic perovskite substrates having a lattice mismatch of -1.6 to 2.3% with reference to bulk $SrRuO_3$. The Poisson ratio, which describes the two orthogonal distortions due to the substrate clamping effect, is estimated to be 0.33. The Curie temperature ($T_C$) and residual resistivity ratios of the series of films are higher than or comparable to the highest reported values for $SrRuO_3$ on each substrate, confirming the high crystalline quality of the films. A $T_C$ of 169 K is achieved in a tensile-strained $SrRuO_3$ film on the $DyScO_3$ (110) substrate, which is the highest value ever reported for $SrRuO_3$. The $T_C$ (146-169 K), magnetic anisotropy (perpendicular or in-plane magnetic easy axis), and metallic conduction (residual resistivity at 2 K of 2.10 - 373 μΩ·cm) of $SrRuO_3$ are widely controlled by epitaxial strain. These results provide guidelines to design $SrRuO_3$-based heterostructures for device applications.

## I. INTRODUCTION

An epitaxially strained heterostructure composed of perovskite-structured oxides offers an exciting platform in which intriguing physical properties, such as magnetism, ferroelectricity, two-dimensional electron gas, and superconductivity emerge.[1-7] The lattice parameters of the epitaxial films differ from those of the bulk due to the epitaxial strain induced by the substrate. Since lattices, electrons, and spins in perovskite oxides are strongly coupled, lattice distortions significantly affect those physical properties.

$SrRuO_3$ (SRO) is an itinerant 4*d* ferromagnetic perovskite with high metallicity, chemical stability, compatibility with other perovskite-structured oxides, and ferromagnetism with strong uniaxial magnetic anisotropy.[8–28] It has been widely used in oxide electronics and spintronics as an epitaxial conducting layer in perovskite heterostructures owing to the above fascinating properties.[15] The physical properties in epitaxial SRO films depend on epitaxial strain, and some of those properties are superior to those of the bulk.[29-40] Examples of such intriguing epitaxial strain effects include perpendicular magnetic anisotropy under compressive strain[14,29,31] and enhancement of the Curie temperature ($T_C$) under tensile strain.[32,33] In particular, the compressive-strain-induced perpendicular magnetic anisotropy is beneficial for scalability and the reduction of power consumption in spintronic devices, such as magnetic random access memory, and thus SRO-based all-oxide spintronic devises have been investigated.[22,23,41] However, control of electrical and magnetic properties with a wide range of epitaxial strain has not yet been systematically investigated, despite its importance for designing heterostructure devices. In addition, the fact that the physical properties of SRO are very sensitive to Ru vacancies and other defects[15,18] has hampered a thorough understanding of the epitaxial strain effect on SRO. Take the ferromagnetic ordering temperature for example: both the enhancement and reduction of $T_C$ in tensile-strained SRO films have been reported.[32,33] Since defects, such as Ru vacancies and orthorhombic domains, are known to reduce $T_C$,[15,18,19,26,27,29,42] it is plausible that these controversial results arise from the difference in the crystalline quality of the films. Thus, it is essential to systematically study the epitaxial strain effect using SRO films with high crystalline quality.

This study presents the electrical and magnetic properties in high-quality SRO films under a wide range of epitaxial strain imposed by cubic or pseudocubic perovskite substrates having a lattice mismatch of -1.6 to 2.3% with bulk SRO. The $T_C$ and residual resistivity ratio (RRR) values of the series of films are higher than or comparable to the highest reported values for SRO on each substrate, confirming the high crystalline quality of the films. The results show that $T_C$ (146-169 K), magnetic anisotropy (perpendicular or in-plane magnetic easy axis), and metallic conduction (residual resistivity at 2 K of 2.10 - 373 μΩ ·cm) of SRO can be widely controlled by epitaxial strain.

## II. EXPERIMENT

We grew high-quality epitaxial SRO films with a thickness of 60 nm in a custom-designed molecular beam epitaxy (MBE) setup equipped with multiple e-beam evaporators for Sr and Ru. To impose a wide range of epitaxial strain on SRO films, we used $(LaAlO_3)_{0.3}(SrAl_{0.5}Ta_{0.5}O_3)_{0.7}$ (LSAT) (001), $SrTiO_3$ (STO) (001), $KTaO_3$ (KTO) (001), $DyScO_3$ (DSO) (110), $TbScO_3$ (TSO) (110), $GdScO_3$ (GSO) (110), $SmScO_3$ (SSO) (110), $NdScO_3$ (NSO) (110), and $PrScO_3$ (PSO) (110) substrates. Rare-earth (*RE*) scandates $RE$ScO$_3$ have the $GdFeO_3$ structure, a distorted perovskite structure with the (110) face corresponding to the pseudocubic (001) face. In-plane lattice constants of these cubic or pseudocubic perovskite substrates $a_{sub}$ and lattice mismatch between the substrates and the bulk SRO are shown in Fig. 1(a). The growth parameters were optimized by Bayesian optimization, a machine learning technique for parameter optimization,[25,43,44] with which we achieved an RRR of 51 for the SRO film on the STO substrate. The growth temperature was 772°C for all substrates. We precisely controlled the elemental fluxes, even for elements with high melting points, e.g., Ru (2250°C), by monitoring the flux rates with an electron-impact-emission-spectroscopy sensor, which were fed back to the power supplies for the e-beam evaporators. The Ru and Sr flux supply rates were 0.365 and 0.980 Å/s, respectively, corresponding to a Ru-rich condition. Excessive Ru is known to be desorbed from the growth surface by forming volatile species such as $RuO_4$ and $RuO_3$ under an oxidizing atmosphere,[11] leading to stoichiometric films. The growth rate of 1.05 Å/s was deduced from the thickness calibration of a thick (63 nm) SRO film using cross-sectional scanning transmission electron microscopy (STEM). This growth rate agrees very well with the value (1.08 Å/s) estimated from the Sr flux rate, confirming the accuracy of the film thickness. During growth, the oxidation was carried out with a mixture of ozone ($O_3$) and $O_2$ gas (~15% $O_3$ + 85% $O_2$), which was introduced at a flow rate of ~2 sccm through an alumina nozzle pointed at the substrate. All SRO films were prepared under the same growth conditions; in particular, the local ozone pressure at the growth surface, being the most important factor in determining the number of Ru vacancies,[18] was precisely kept constant. Further information about the MBE setup and preparation of the substrates is described elsewhere.[45–47] X-ray diffraction (XRD) measurements were performed with a Bruker D8 diffractometer using monochromatic Cu $K_{\alpha 1}$ radiation at room temperature. Electrical resistivity was measured in a Physical Property Measurement System (PPMS) using a standard four-probe method with Ag electrodes deposited on the SRO surface without any additional processing. The distance between the two voltage electrodes was 2 mm. The magnetization was measured with a Quantum Design MPMS3 SQUID-VSM magnetometer.

## III. RESULTS AND DISCUSSION

### A. Crystallographic analyses

We characterized the crystallographic properties of the SRO films by $\theta$–$2\theta$ XRD and high-resolution X-ray reciprocal space mapping (HRRSM). Figure 1(b) shows the out-of-plane $\theta$-$2\theta$ XRD patterns around the (002) pseudocubic diffractions of the SRO films, hereafter called $(002)_{pc}$. The SRO $(002)_{pc}$ peaks are located at the lower angle side of the LSAT and STO (002) peaks and the higher angle side of the DSO, TSO, GSO, KTO, SSO, NSO, and PSO (002) or $(002)_{pc}$ peaks, indicating the $RuO_6$ octahedra are vertically elongated and compressed by the substrate-induced in-plane compressive and tensile strain, respectively. The out-of-plane lattice constants $c_{film}$ estimated from the Nelson-Riley extrapolation method for $\theta$-$2\theta$ XRD patterns of the films are listed in Table I. In the $\theta$-$2\theta$ XRD patterns for the SRO films on the STO, DSO, TSO, GSO, and SSO substrates, Laue fringes are clearly observed [Fig. 1(c)]. The periods of the Laue fringes in these films are almost identical, as indicated by the dashed lines in Fig. 1(c). The film thicknesses estimated from the periods of the Laue fringes (63 nm) agree very well with those determined by STEM (60 nm), reconfirming the accuracy of the thickness in addition to the high crystalline quality and large coherent volume of the films. On the other hand, the broad $(002)_{pc}$ peaks for the SRO films on the LSAT, KTO, NSO, and PSO substrates are not accompanied by Laue fringes, indicating that they are not uniform due to the large lattice mismatch [Fig. 1(a)].

Figure 2 shows the HRRSM around the (103) pseudocubic diffractions of the SRO films, hereafter called $(103)_{pc}$. In the SRO films on the STO [Fig. 2(b)], DSO [Fig. 2(c)], TSO [Fig. 2(d)], GSO [Fig. 2(e)], and SSO [Fig. 2(g)] substrates, horizontal peak positions of SRO are identical to those of the substrates, confirming that these films were grown coherently. However, coherent growth was not maintained for SRO films grown on the LSAT [Fig. 2(a)], KTO [Fig. 2(f)], NSO [Fig. 2(h)], and PSO [Fig. 2(i)] substrates, which is consistent with the $\theta$-$2\theta$ XRD patterns. The striking contrast between the films on SSO ($a_{dub}$ = 3.990 Å) and KTO ($a_{dub}$ = 3.989 Å) cannot be solely attributed to the difference in the lattice constant of the substrates. Instead, it might have stemmed from the difference in the crystal structure (the $GdFeO_3$ structure for SSO; the perovskite structure for KTO) and/or surface conditions of the substrates. The in-plane lattice constants $a_{film}$ estimated from the SRO $(103)_{pc}$ peaks are listed in Table I.

From the $a_{film}$ and $c_{film}$ values listed in Table I, we calculated the unit-cell volumes $V_{film} = a_{film}^2 c_{film}$ of the SRO films. Figure 3(a) shows the $V_{film}$ of the SRO films plotted as a function of the $a_{film}$. The $V_{film}$ increases linearly with strain as predicted by Zayak *et al.*[48] Our experimentally determined absolute volume values are higher than those predicted by local spin-density approximation (LSDA) methods, probably because LSDA underestimates the lattice constant values.[48] To analyze the relation between lateral and perpendicular strain, we plotted the in-plane strain $\varepsilon_a$ of the SRO films as a function of the out-of-plane strain $\varepsilon_c$ [Fig. 3(b)]. The Poisson ratio of epitaxial films describes the two orthogonal distortions due to the clamping effect of the substrates, which can be expressed as $\nu = \varepsilon_c/(\varepsilon_c - 2\varepsilon_a)$,[49,50] leading to the relationship $\varepsilon_a/\varepsilon_c = (\nu - 1)/(2\nu)$. The Poisson ratio of the SRO films obtained by the linear fitting of the data is 0.33 [Fig. 3(b)], which is typical of many non-ferroelectric perovskites.[49,50] This value also coincides with results from density functional theory calculations within the LSDA (LSDA-DFT), from which a Poisson ratio of $\nu = 0.32$ can be estimated.[48]

**B. Electrical and magnetic properties**

Figure 4(a) shows a logarithmic plot of the temperature dependence of the resistivity $\rho$ for the SRO films. For all SRO films, a clear kink at 146-169 K is observed. These kinks can be more clearly seen in the normalized resistivity vs. temperature curves [Fig. 4(b)]. The kinks correspond to the $T_C$ at which the ferromagnetic transition occurs and below which spin-dependent scattering is suppressed.[14,15] The $T_C$, RRR, and the residual resistivity at 2 K [$\rho$(2K)] largely vary depending on the degree of strain. As shown in Table I, the $T_C$ and RRR values determined in this study are higher than or comparable to the highest reported values for SRO on each substrate. The outstanding crystalline quality of the films allows us to access intrinsic properties of SRO since defects, such as Ru vacancies and orthorhombic domains, are known to reduce $T_C$ and RRR.[15,18,19,27,29,42] The controllability of the magnetic and electrical properties by strain demonstrated here should provide new insights into strain-driven electron correlations.

To clarify the correlation between the epitaxial strain and ferromagnetic order, we plotted $T_C$ of the SRO films as a function of the $a_{film}$ [Fig. 5(a)]. The compressive-strained SRO films have lower $T_C$, while the tensile-strained ones have higher $T_C$ than the bulk SRO (160 K). Particularly, the $T_C$ value of the tensile-strained SRO film on the DSO substrate (169 K, $a_{film}$ = 3.942 Å) is the highest value ever reported for SRO.[15] The coherently grown tensile-strained SRO films have higher $T_C$ than those of the relaxed tensile-strained SRO films. This result means that the misfit dislocations,[51] which relax the epitaxial strain at the interface, are responsible for a suppression of the ferromagnetic order. The strain-induced enhancement and reduction of the ferromagnetic order may originate from the change in Ru–O–Ru bond angles and/or Ru–O bond lengths, which affects the overlapping of Ru $t_{2g}$ and O $2p$ orbitals.[29,48,52] In fact, LSDA-DFT calculations predict the enhancement of the ferromagnetic order through the change in the Ru–O bond lengths[52] under tensile strain in SRO films. That is, the $T_C$ gradually increases with increasing tensile strain, and it becomes maximum when $a_{film}$ = 3.970 Å. However, the experimental $a_{film}$ value at which $T_C$ is highest is smaller than the calculated value. Besides, for compressive strain, enhancement of $T_C$ is also expected by LSDA methods.[52] Thus, our results will motivate further theoretical investigation of the ferromagnetic order in epitaxially strained SRO films.

Figure 5(b) shows the residual resistivity $\rho$(2 K) of the SRO films plotted as a function of $a_{film}$. The low $\rho$(2 K) of the SRO films on the STO (2.10 μΩ·cm) and DSO (2.14 μΩ·cm) substrates accentuate the low concentration of defects, which contributes to the high RRR of 51 and 72, respectively. Importantly, the RRR value of the SRO film on the DSO substrate is higher than that on the STO substrate, meaning that the SRO film on the DSO substrate has fewer defects owing to the smaller lattice mismatch. Thus, the SRO/DSO interface may overcome the interface-driven disorder problem that degrades magnetic and metallic properties of SRO near heterointerfaces.[27,53,54] The $\rho$(2 K) increases with decreasing (increasing) $a_{film}$ of the compressive-strained (tensile-strained) SRO films, indicating that the number of defects increases with increasing epitaxial strain. Notably, even though the compressive strain of the SRO film on the LSAT substrate is considerable (1.3%), it has a lower $\rho$(2 K) (5.72 μΩ·cm) than those of the tensile-strained SRO films, except for the SRO film on the DSO substrate. The observations indicate that the metallic conduction in SRO is robust under compressive strain. Since compressive-strained SRO films have a larger tolerance for good metallic conduction, they are more suitable as a metallic oxide electrode, which can lead to lower Joule heating and energy loss in heteroepitaxially grown devices.[55]

Figure 6 plots the magnetization-versus-magnetic field curves of the SRO films on the STO (compressive) and KTO (tensile) substrates at 10 K with a magnetic field applied to the out-of-plane [001] or in-plane [100] direction of the substrates. The saturation magnetization of the compressive-strained and tensile-strained SRO films along the out-of-plane [001] and in-plane [100] directions, respectively, is 1.25 $\mu_B$/Ru, which is a typical value for bulk and thin-film SRO.[15] In contrast, the magnetization of the compressive-strained and tensile-strained SRO films along the in-plane [100] and out-of-plane [001] directions, respectively, at 7 T is lower than saturated value (1.25 $\mu_B$/Ru). These results indicate that the easy magnetization directions are perpendicular and in-plane for the compressive-strained and tensile-strained SRO films, respectively, as is the case with compressive-strained and tensile-strained SRO films reported previously.[15] Due to the large crystal field splitting, $Ru^{4+}$ states in SRO are expected to be in a low-spin state with an electron configuration of $t_{2g}^4$ (3↑, 1↓) corresponding to 2 $\mu_B$/$Ru^{4+}$ ion. The lower experimental saturation magnetization in SRO has been attributed to electron delocalization associated with itinerancy.[56,57]

## IV. SUMMARY

We have investigated the epitaxial strain effect on the electrical and magnetic properties of high-quality SRO films. A wide range of epitaxial strain was imposed on SRO films by using various cubic perovskite and pseudocubic rare-earth scandate substrates. The $T_C$ and RRR values obtained in this study are higher than or comparable to the highest reported values for SRO on each substrate, confirming the overall superior crystalline quality of SRO films grown by MBE. $\theta$–$2\theta$ XRD and HRRSM measurements revealed that the SRO films on STO, DSO, TSO, GSO, and SSO substrates are grown coherently. In contrast, the SRO films on LSAT, KTO, NSO, and PSO substrates are relaxed due to the large lattice mismatch. The $T_C$ (146-169 K), magnetic anisotropy (perpendicular or in-plane magnetic easy axis), metallic conduction [$\rho$(2 K) = 2.10-373 μΩ · cm] of SRO were widely controlled by epitaxial strain. The stability of the ferromagnetic order in SRO was enhanced by tensile strain, and the $T_C$ became maximum when $a_{film}$ = 3.942 Å (on DSO, 169 K). The Poisson ratio obtained by the linear fitting of the $\varepsilon_a$-$\varepsilon_c$ relationships was 0.33, which is a typical value of many non-ferroelectric perovskites.[49,50] The 1.3% compressive-strained SRO film on the LSAT substrate has low $\rho$(2 K) (5.72 μΩ ·cm), indicating the metallic conduction in SRO is more robust under compressive strain. Our results provide important insights into the epitaxial strain effect of SRO and for applying SRO as a metallic ferromagnetic oxide electrode for hetero-epitaxially grown electrical and spintronic devices.

**AUTHORS' CONTRIBUTIONS**

Y.K.W. conceived the idea, designed the experiments, and led the project. Y.K.W. and Y.K. grew the samples. Y.K.W. carried out the X-ray diffraction and magnetic measurements. S.K.T. and Y.K.W. carried out the transport measurements. Y.K.W. analyzed and interpreted the data. All authors contributed to the discussion of the data. Y.K.W. wrote the paper with input from all authors.

**DATA AVAILABILITY**

Data that support the findings of this study are available from the corresponding author upon reasonable request.

## Figures and figure captions

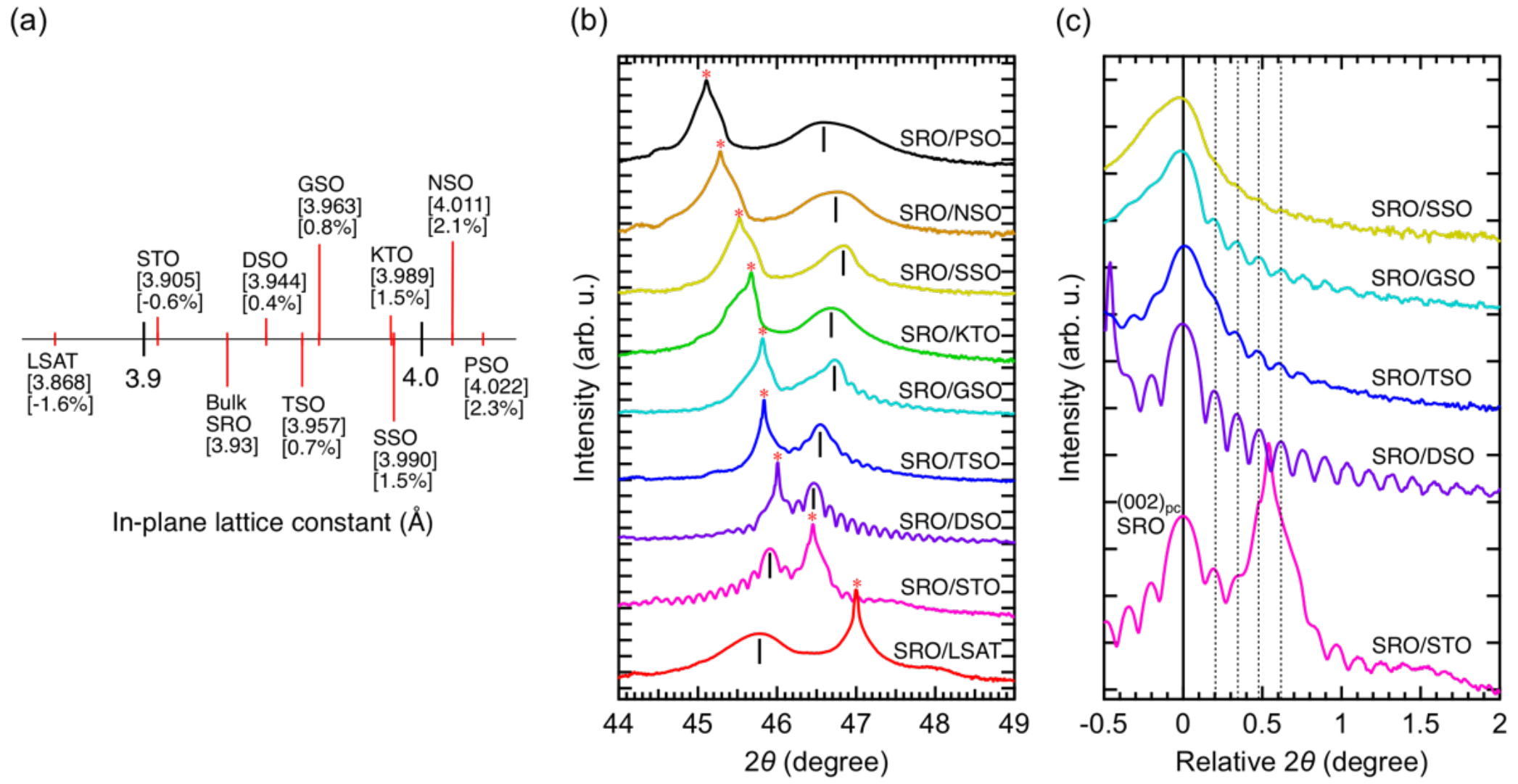

FIG. 1. (a) In-plane lattice constants of the cubic or pseudocubic perovskite substrates and lattice mismatch between the substrates and the bulk SRO. (b) Out-of-plane $\theta$-$2\theta$ XRD patterns around the $(002)_{pc}$ diffractions of the SRO films on the various substrates. The black lines and asterisks indicate the peak positions of the $(002)_{pc}$ diffractions of the SRO films and the peak positions of the (002) or $(002)_{pc}$ diffractions of the substrates, respectively. (c) Out-of-plane $\theta$-$2\theta$ XRD patterns plotted as a function of the relative $2\theta$ to the $(002)_{pc}$ peaks of the coherently grown SRO films. Dashed lines indicate the peak positions in the Laue fringes for SRO/DSO.

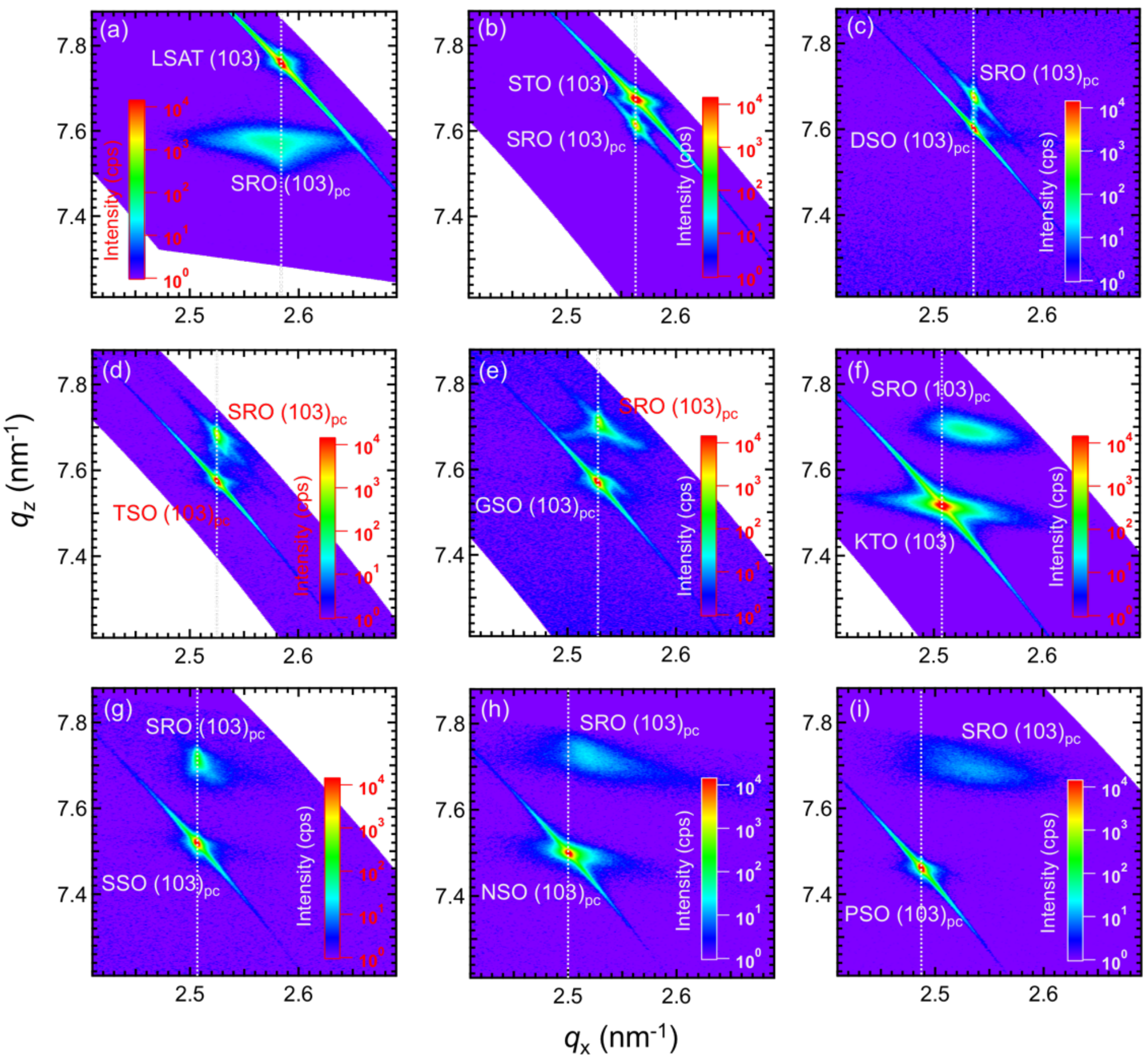

FIG. 2. X-ray HRRSM around the $(103)_{pc}$ diffractions of the SRO films on LSAT (a), STO (b), DSO (c), TSO (d), GSO (e), KTO (f), SSO (g), NSO (h), and PSO (i) substrates. The white dashed lines indicate the peak positions of the (103) or $(103)_{pc}$ diffractions of the substrates.

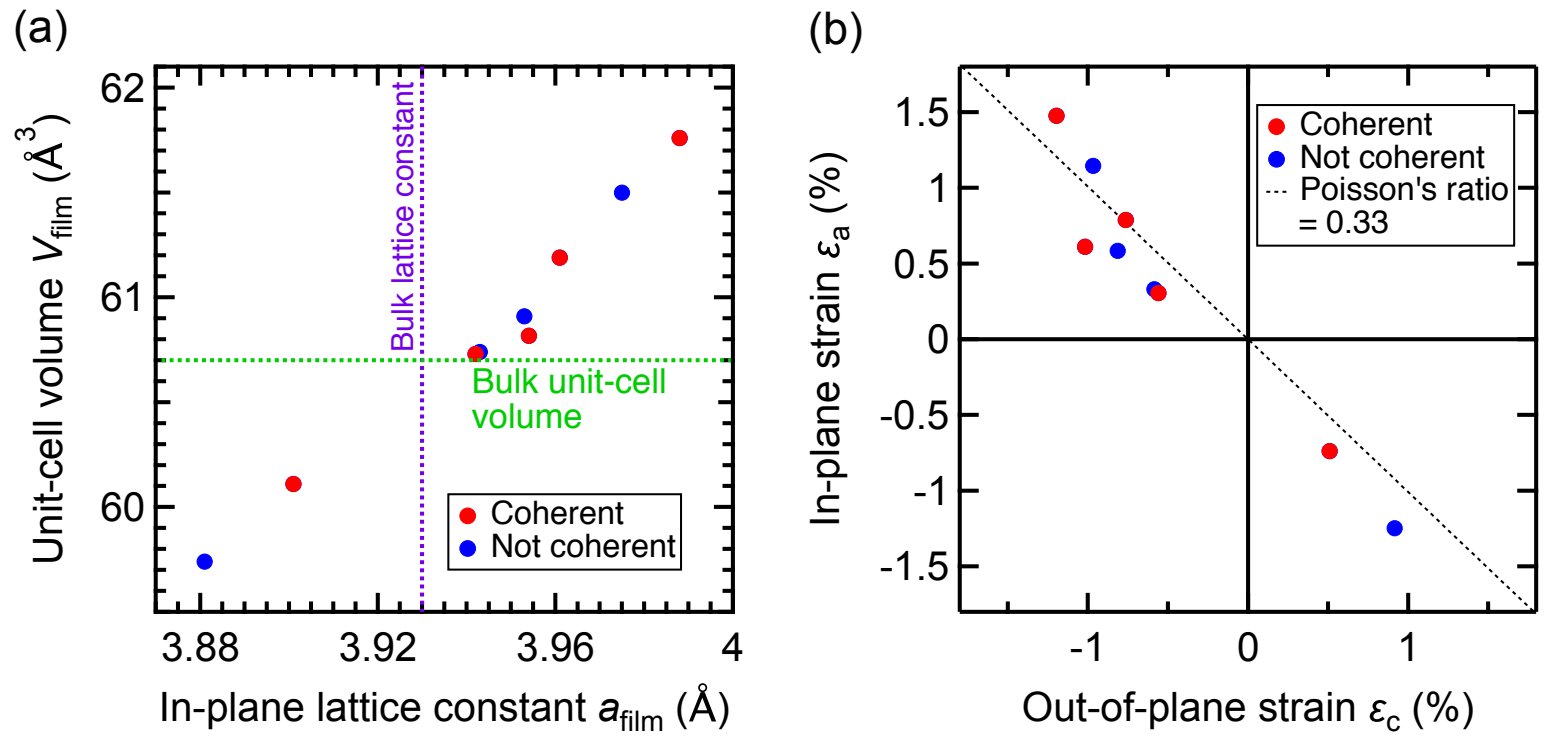

FIG. 3. (a) Unit-cell volume of the SRO films plotted as a function of the in-plane lattice constant $a_{film}$. (b) In-plane strain $\varepsilon_a$ of the SRO films plotted as a function of the out-of-plane strain $\varepsilon_c$. The dashed line corresponds to Poisson's ratio of 0.33.

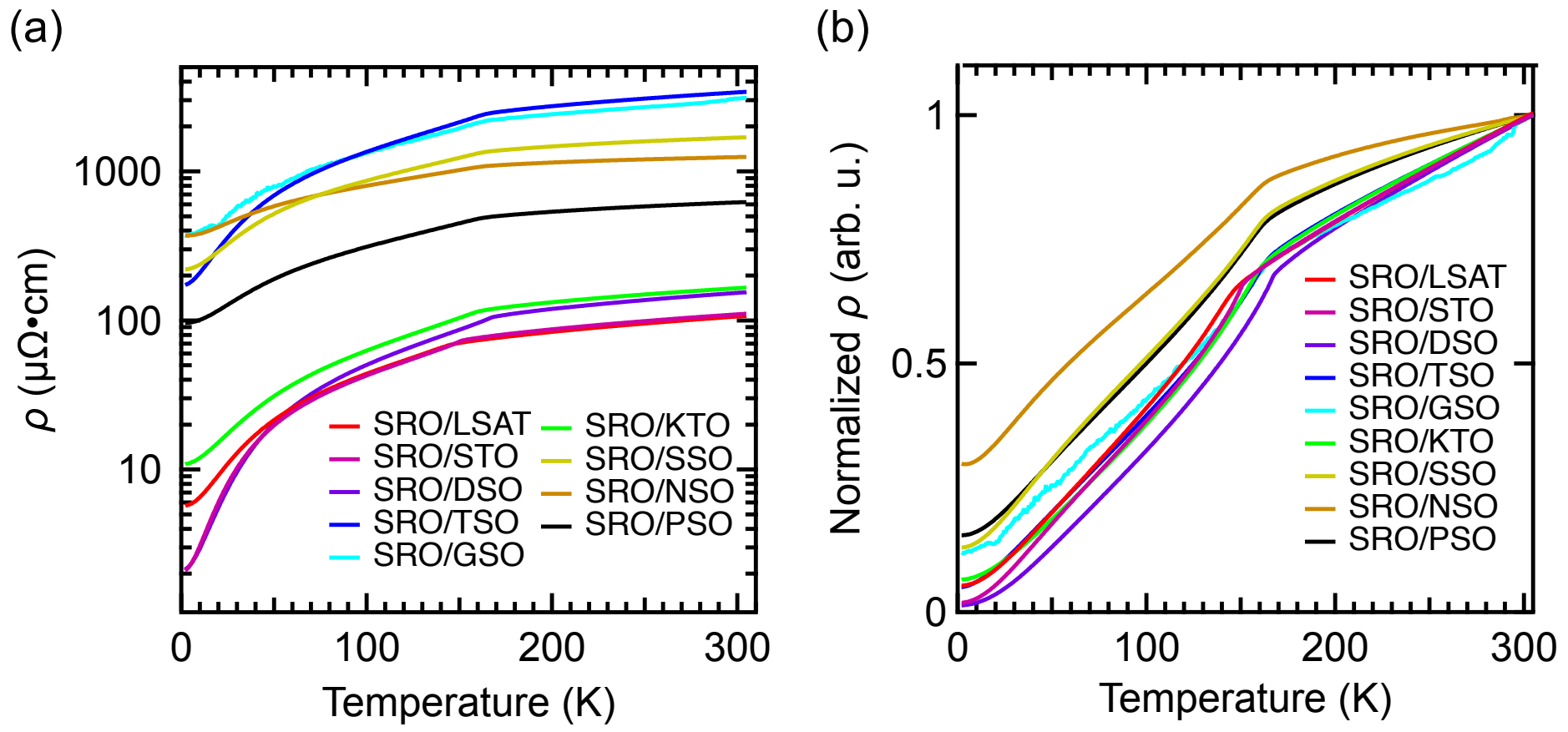

FIG. 4. (a) Logarithm plot of the temperature dependence of the resistivity $\rho$ for the SRO films on the various substrates. (b) Temperature dependence of the $\rho$ normalized at 300 K for the SRO films on the different substrates.

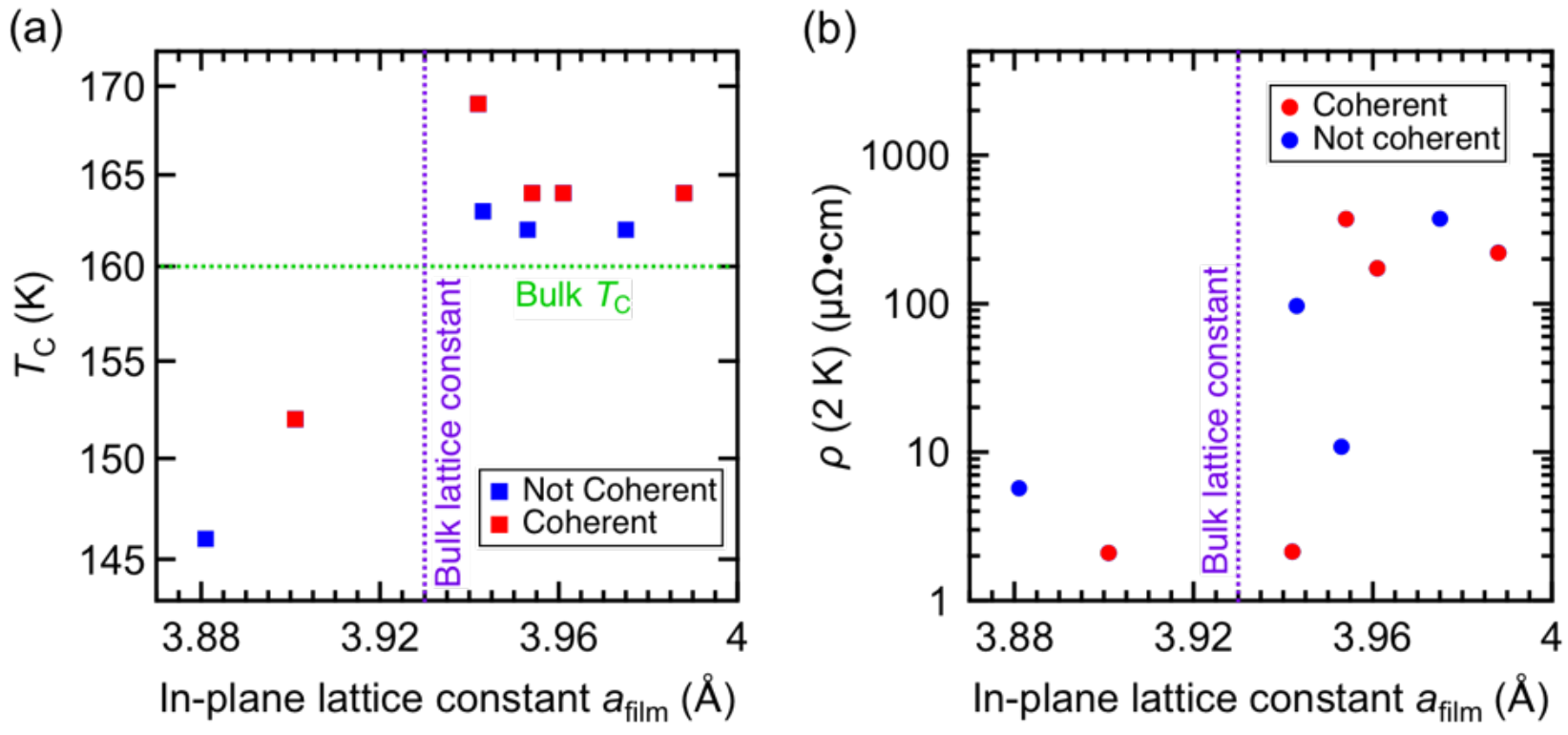

FIG. 5. (a) $T_C$ and (b) residual resistivity $\rho$(2 K) of the SRO films plotted as a function of the in-plane lattice constant $a_{film}$.

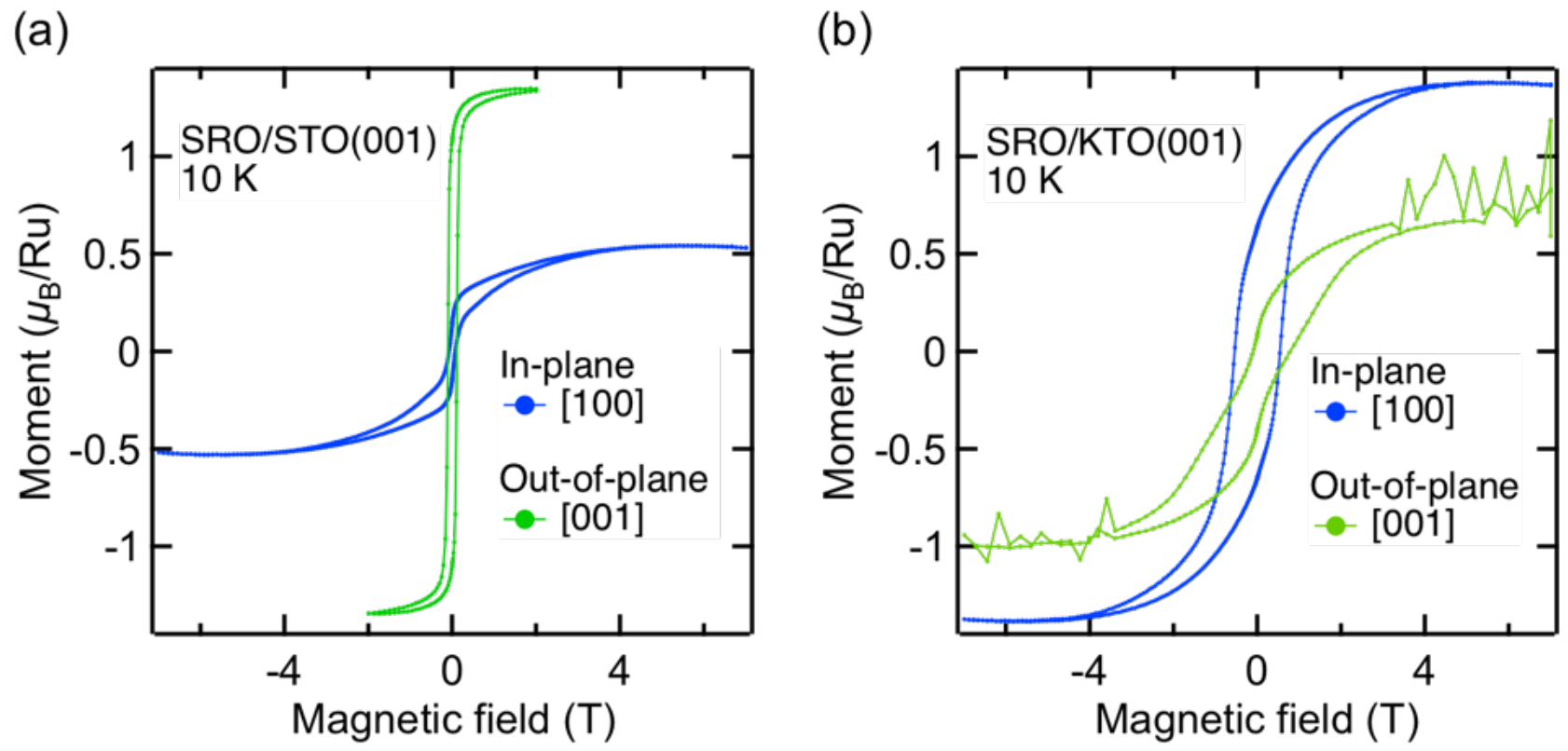

FIG. 6. Magnetization vs. magnetic field curves at 10 K for the SRO films on the STO (a) and KTO (b) substrates. The magnetic fields were applied to the out-of-plane [001] (green circles) and in-plane [100] (blue circles) directions of the substrates.

**TABLE I.** Physical properties of the SRO films grown on the cubic or pseudocubic perovskite substrates. The $T_C$ and RRR values reported in previous studies are also listed.

| Substrate | $T_C$ (K) | $T_C$ (K) Previous studies | RRR | RRR Previous studies | $\rho$(2K) (μΩ·cm) | $c_{film}$ (Å) | $a_{film}$ (Å) | $a_{sub}$ (Å) | Growth |
|---|---|---|---|---|---|---|---|---|---|
| LSAT(001) (comp.) | 146 | 142[36] | 19 | 4[36] | 5.72 | 3.966 | 3.881 | 3.868 | not coherent |
| STO(001) (comp.) | 152 | 130-152[14,26,57] | 51 | 2-84[14,26,57] | 2.10 | 3.950 | 3.901 | 3.905 | coherent |
| DSO(110) (tensile) | 169 | 140-168[32,35] | 72 | 3-74[32,35] | 2.14 | 3.908 | 3.942 | 3.944 | coherent |
| TSO(110) (tensile) | 164 | No report | 20 | No report | 173 | 3.900 | 3.961 | 3.957 | coherent |
| GSO(110) (tensile) | 164 | 100-163[19,34,37] | 8 | 2-5[19,34,37] | 372 | 3.890 | 3.954 | 3.963 | coherent |
| KTO(001) (tensile) | 162 | 150[39] | 15 | 3[39] | 10.9 | 3.898 | 3.953 | 3.989 | not coherent |
| SSO(110) (tensile) | 164 | 125-128[40] | 8 | 2[40] | 220 | 3.883 | 3.988 | 3.990 | coherent |
| NSO(110) (tensile) | 162 | No report | 3 | No report | 373 | 3.892 | 3.975 | 4.011 | not coherent |
| PSO(110) (tensile) | 163 | No report | 6 | No report | 96.6 | 3.907 | 3.943 | 4.022 | not coherent |